\documentstyle[aps,prb,12pt]{revtex}
\begin{document}
\tightenlines
\def\beq{\begin{equation}}
\def\eeq{\end{equation}}
\def\bea{\begin{eqnarray}}
\def\eea{\end{eqnarray}}
\def\ve{\vert}
\def\vel{\left|}
\def\ver{\right|}
\def\nnb{\nonumber}
\def\ga{\left(}
\def\dr{\right)}
\def\aga{\left\{}
\def\adr{\right\}}
\def\rar{\rightarrow}
\def\nnb{\nonumber}
\def\la{\langle}
\def\ra{\rangle}
\def\ba{\begin{array}}
\def\ea{\end{array}}
\def\ds{\displaystyle}

\def\R{\psi^R}
\def\L{\psi^L}
\def\br{\ve \la}
\def\ke{ \ra \ve^2}
\def\dz{\ve \partial_z \ve}
\def\H{{\cal H}}
\def\S{{\cal S}}
\def\k{{\bf k}}
\def\w{\omega}
\def\bos{\lower 0.5cm\hbox{{\vrule width 0pt height 1.3cm}}}
\def\aaa{\lower 0.cm\hbox{{\vrule width 0pt height .8cm}}}
\def\dol{\lower 0.6cm\hbox{{\vrule width 0pt height .8cm}}}
\baselineskip  0.7cm

\title{
Reduced Density Matrix Approach to Phononic Dissipation in Friction} 

\author{{
A. \"{O}ZP\.{I}NEC\.{I}$^\dagger$, \,
D. LEITNER$^*$, \,
S. {\c C}IRACI$^\dagger$} \\ 
{$^\dagger$ Department of Physics, Bilkent University 06533 Ankara, 
Turkey \\} 
{$^*$ Department of Chemistry, University of California, San Diego, 
CA} } 
\date{\today}

\begin{titlepage}
\maketitle
\begin{abstract}
Understanding mechanisms for energy dissipation from nanoparticles 
in contact with large samples is a central problem in describing
friction microscopically.  Calculation of the reduced 
density matrix appears to be the most suitable method to study such 
systems that are coupled to a large environment.  In this paper, the time 
evolution of the reduced density matrix has been evaluated for an 
arbitrary system coupled to a heat reservoir.  The formalism is then applied 
to study the vibrational relaxation following the
stick-slip motion of a small adsorbate on a surface.
The frequency dependence of the relaxation time is also determined.\\
PACS: 44.10.+i;62.20.Qp;63.22.+m;67.55.Hc
\end{abstract}

%\pacs{PACS: 44.10.+i;62.20.Qp;63.22.+m;67.55.Hc}
\end{titlepage}

\section{Introduction}
Friction and its microscopic origins have been intensively investigated
in recent years.\cite{genfric}  Progress in atomic force microscopy (AFM) has made
possible precise force measurements at the atomic scale.\cite{mmrc87}  These data,
together with simulations, theory and {\it ab initio} force calculations
have helped to provide a detailed picture of friction.
\cite{bil95,sp90,sjs96,fs99,zuk99,csr94,js90,js93,tswk97,bc97,blc99}  
Much of this work is concerned with the nature and rate of energy transfer from
lubricant layers or asperities into bulk substrates.  The understanding
of such processes has implications as far-ranging as the design
of solar collectors,\cite{solar} where collected energy must be transferred
to the bulk before reemission to the surroundings; to the understanding
of reactions in living cells catalyzed on small surfaces,\cite{catalysis} as well
as biomolecular motors;\cite{rk95} and to the design of lubricants for industrial
purposes.\cite{rk95}  In many cases, friction is the result of energy dissipation by
multiphonon processes.  We address in this article energy dissipation
in a stick-slip model for friction involving small and sparsely distributed
adsorbates on a surface.  We determine the nature and calculate the
rate of damping of the adsorbed molecule, nanoparticle or asperity
following the slip step.

Relaxation processes of adsorbates have been studied both theoretically 
and experimentally (see eg. 
\cite{persson2,persson3,persson4,hoffmann,persson5}).  In most of 
these works, the authors used either phenomenological approximations 
\cite{persson4} or 
assumed that the system under study is at least at quasi-equilibrium and 
used equilibrium theory to study their properties 
\cite{persson2,persson3,persson5,persson1}.
In this article, a Redfield Theory-like approach is developed 
(for the derivation of the Redfield theory and some of its applications 
see, e.g., Refs. [\cite{redfield,levy,blum}]) to calculate the time evolution
of the reduced density matrix of an adsorbate on a substrate within
the framework of a stick-slip model for friction.  

Consider an object adsorbed on the surface of a sample, a metal or
insulator, with vibrational
frequency $\Omega$. In general there are two possible decay modes: 1) it can 
create electronic excitations in a metal, eg., create electron-hole 
pairs, or 2) it can create phononic excitations.  In this article our 
interest will be in phononic dissipation.  If $\Omega \simeq n\, \omega_0$ 
where $\omega_0$ is the maximum phonon frequency of the sample, the excitations
can decay only by the creation of $n$ phonons in the sample 
\cite{persson4}.  For large $n$, 
this contribution is in general negligible.  Thus for small particles on
a solid substrate, for which the vibrational frequency of the adsorbate
might be much higher than the band of substrate mode frequencies, energy
dissipation from the adsorbate into the substrate may be quite slow.
For such cases, anharmonic coupling between the adsorbate and substrate
gives rise to energy flow into the substrate.   For example, for CO
adsorbed on Cu, the
Cu-CO stretch vibration is $\Omega \simeq 1.5 \, \omega_0$ and decay by creating 
two phonons might be an important mechanism for the vibrational damping 
of the molecule.  We will address the problem of phononic energy
dissipation in friction following stick-slip motion of
an adsorbate whose vibrational frequency
lies above the band of substrate frequencies, like CO on a Cu surface.
In Ref. [\cite{persson1}]  two and three phonon contributions to the 
dissipation of the Cu-CO stretch vibrations was studied using the Golden Rule 
formula.  We will 
study this and similar systems here using a more general Redfield 
theory-like approach for various $\Omega$'s to understand the dependence 
of the dissipation rate on the coupling between the sample and the 
adsorbed atom.  
We assume that the adsorbates are sparsely distributed so that
we can neglect interactions among them.

The organization of the paper is as follows: In Sec. 
II, we calculate the time evolution of 
the reduced density matrix which allows one to take into account all 
non-equilibrium properties of a system and also takes into account 
possible coherence and incoherence effects (for the properties of density 
matrices see, e.g. Ref. [\cite{blum}]). Possible limitations on the obtained 
evolution are also discussed.  In Sec. III a model system is proposed 
which is analyzed and solved in  Sec. IV.  Concluding remarks are
given in Sec. V.

\section{Evolution of the Reduced Density Matrix}
In studying the dynamics of systems coupled to an environment, it is most 
natural to use the Reduced Density Matrix (RDM) 
formalism. The reduced density matrix, $\rho_R$, is  obtained from the 
density  matrix, $\rho$, of  the system plus the environment by taking its 
trace over the environmental degrees of freedom,
\bea
\rho_R(t) = Tr_b\rho(t) \, , \label{reduced}
\eea
where $Tr_b$ denotes trace over the degrees of freedom of the 
environment. In the 
following, the index $R$ is omitted whenever there is no possibility of 
confusion.  Once the  time evolution of the  RDM of the system is known, 
the  time evolution of the expectation value of any observable, ${\cal 
O}$, of the system can be obtained as:
\bea
\la {\cal O} \ra (t) = Tr[\rho_R (t) {\cal O}] \, ,
\eea

The time dependence of the RDM of the system can be obtained 
from the definition of the RDM in Eq. (\ref{reduced}).
Let
\bea
\H = \H_s + \H_r + \H_{int} \, ,
\eea
where $\H_s$, $\H_r$ are the system and reservoir Hamiltonians, 
respectively, and $\H_{int}$ describes the interaction between them.  We 
will assume, without loss of generality, that
\bea
\H_{int} = \sum_s Q_s F_s
\eea
where $Q_s(F_s)$ acts only on the system (reservoir) degrees of freedom.
The time evolution of the components of the RDM is given by
\bea
\rho_{\alpha \beta} (t) &=& \rho_{\alpha \beta}(0) e^{-i \omega_{\alpha
\beta} t } + \sum_{\alpha' \beta'} R_{\alpha \alpha' ; \beta \beta'} (t)
\rho_{\alpha' \beta'}(0) e^{-i \omega_{\alpha \beta} t } \, ,
\eea   
where $\hbar \omega_{\alpha \beta} = \epsilon_\alpha - \epsilon_\beta$ and
 the tensor $R_{\alpha \alpha' ; \beta \beta'} (t)$ is defined as
\bea   
R_{\alpha \alpha' ; \beta \beta'} (t) = \sum_{ij}  P(E_j) \la \alpha i
\ve \S (t) \ve \alpha' j \ra \la \beta' j \ve \S^\dagger (t) \ve \beta i \ra
-\delta_{\alpha \alpha'} \delta_{\beta \beta'},
\eea
where the scattering matrix, $\S (t)$, is defined as
\bea
\S (t) &=& e^{\frac{i}{\hbar} \H_0 t} e^{\frac{-i}{\hbar} \H t} \nnb \\
&=& 1 - \frac{i}{\hbar} \int_0^t dt' \H_{int}(t') +
\left(- \frac{i}{\hbar}\right)^2
\int_0^t dt_1 \int_0^{t_1} dt_2 \H_{int}(t_1) \H_{int}(t_2)+ \, ...
\label{smatrix} 
\eea
Here, $\H_0$ is defined to be $\H_0 = \H_r +\H_s$. Also
$\H_{int}(t) = e^{\frac{i}{\hbar} \H_0 t} \H_{int} e^{-\frac{i}{\hbar} \H_0
t} \, ,
$ and  $\ve \gamma j \ra = \ve \gamma \ra \otimes \ve j \ra$ with,
\bea
\H_{s} \ve \gamma \ra &=& \epsilon_\gamma \ve \gamma \ra \nnb \\
\H_r \ve j \ra &=& E_j \ve j \ra \, .
\eea
In the following Greek (Latin) letters will denote the system (reservoir)
degrees of freedom.
In deriving this result it is 
assumed that the reservoir is always in equilibrium so that the density  
matrix of the whole system can be factorized as
\bea
\la \gamma j \ve \rho(t) \ve \delta k \ra =\delta_{jk} \, \rho_{R\gamma 
\delta} \, P(E_j)
\eea
where the diagonal density matrix elements of the reservoir are defined as
\bea
P(E_j) = \frac{e^{-\beta E_j}}{Z} \, ,
\eea
where $Z = \sum_j e^{- \beta E_j}$.

%\bea
%\S (t) &=& e^{\frac{i}{\hbar} H_0 t} e^{\frac{-i}{\hbar} H t} \nnb \\
%&=& 1 - \frac{i}{\hbar} \int_0^t dt' \H_{int}(t') + 
%\left(- \frac{i}{\hbar}\right)^2 
%\int_0^t dt_1 \int_0^{t_1} dt_2 \H_{int}(t_1) \H_{int}(t_2)+ \, ... 
%\label{smatrix}
% \eea 

Until this point, the only assumption made is that the density matrix of 
the whole system is factorizable which resulted in a linear equation for the 
components of the RDM.  The applicability of this approximation should be 
studied carefully.  This assumption is valid only if there exists a weak 
coupling between the system and the reservoir so that the tensor product 
states $\ve \alpha j \ra$ can be considered to be nearly the eigenstates of 
the whole system.  If there is strong coupling between the system and 
the reservoir, or if the ``reservoir" is a finite one, the density matrix 
of the whole system in general cannot be factorized and one has to do 
without this simplifying approximation.  This approximation should be valid
for our study of phononic energy decay from an adsorbate coupled
anharmonically to a large substrate.

Now, the main task is to find a suitable approximation for the tensor 
$R_{\alpha \alpha' ; \beta \beta'} (t)$. Once it is known, the time 
evolution of the RDM can be calculated.
Unfortunately, the expression obtained by 
straightforward application of 
the second order expansion of the $\S$ matrix yields a result which is 
valid only if the time, $t$, is sufficiently short.
To overcome this difficulty we used an 
iterative scheme in which we calculated the initial RDM and then evolved 
it one step in time; taking the evolved RDM as the initial RDM, we 
evolved it one step further.  At each step, we evolved for a short 
enough time.   
Since energy is not conserved for finite times, one has to impose 
energy conservation by hand.  For this reason, 
the matrix elements of $\H_{int}$ coupling states of different energies 
are neglected. The 
calculations are similar to the ones done in scattering theory with the 
result 
\bea
\rho_{\alpha \beta}(t+\Delta t) &=& \rho_{\alpha \beta}(t) e^{-i 
\omega_{\alpha \beta} \Delta t} + \sum_{\alpha' \beta'} R_{\alpha 
\alpha'; \beta \beta'} (\Delta t) \rho_{\alpha' \beta'}(t) e^{-i 
\omega_{\alpha \beta} \Delta t} \, ,\\
R_{\alpha \alpha' ; \beta \beta'}( \Delta t) &=& \frac{ \Delta t}{\hbar^2} 
\delta_{\alpha - \alpha',\beta - \beta'} \left( \sum_{s s'} Q_s^{\alpha 
\alpha'} Q_{s'}^{\beta' \beta} j_{s s'} (\omega_{\alpha \alpha'}) - \right. 
\nnb \\  
&-& \frac{1}{2} \delta_{\alpha \alpha'} \sum_{ss'} \sum_\gamma Q_s^{\beta 
\gamma} Q_{s'}^{\gamma \beta} j_{s s'} (\omega_{\gamma \beta}) - \nnb \\ 
&-&\left. \frac{1}{2}  \delta_{\alpha \alpha'} \sum_{ss'} \sum_\gamma 
Q_s^{\alpha \gamma} Q_{s'}^{\gamma \alpha} j_{s s'} (\omega_{\gamma 
\alpha}) \right) \, , \\
j_{ss'}(\omega) &=& \Delta t \sum_{kj} ' P(E_j) F_s^{kj} F_{s'}^{jk}\, , 
\label{last}
\eea
where the prime on the summation in Eq. (\ref{last}) indicates that the sum 
should be 
carried out over states for which $\hbar \omega = E_{jk}$.  
  
\section{The Model Hamiltonian}
Consider an object adsorbed on a surface.  Let $M$ be the mass of a 
reservoir atom and $m$ be the mass of the adsorbed object.  Assume that 
the small adsorbate is bonded to a single atom of the sample and the  
interaction between the sample atom and the adsorbed atom is described by 
the Morse potential,
\bea
U(u-v) = E_0 \left\{ e^{-2 \alpha (u-v)} - 2 e^{- \alpha (u-v)} \right\} 
\, , 
\eea
where $u$ and $v$ are the vertical displacements of the adsorbed atom and 
the sample atom, respectively. $E_0$ is the binding energy of the adsorbed 
atom and $\alpha$ can be related to the vibrational frequency, $\Omega$, of 
the adsorbed atom through
\bea
\alpha = \left( \frac{m \Omega^2}{2 E_0} \right)^\frac{1}{2} \, .
\eea 
Expanding the potential and retaining the lowest order terms, we get
\bea
\H_{int} = A u v + B u v^2
\eea
where 
\bea
A &=& -2 E_0 \alpha^2  \\
B &=& -3 E_0 \alpha^3.
\eea
For $\Omega > \omega_0$, the $uv$-term makes no contribution to dissipation since it 
does not conserve energy, and we only have the $uv^2$-term. For the other 
case, $\Omega < \omega_0$, the contribution to the decay rate from
the $uv$ term is generally much larger than from the $uv^2$ 
term.  The decay of the vibrational excitation of the adsorbate to
the substrate for the case where $\Omega < \omega_0$ using only
harmonic coupling has been studied by exact
diagonalization of the Hamiltonian \cite{exact}. The calculated value 
for the decay rate in Ref. [\cite{exact}] is two orders greater than the value we have 
calculated in Sec. IV, where we used only the $uv^2$ term.  
In this article, we consider only effects of the $uv^2$ term.  We thus
focus on energy dissipation for high-frequency adsorbates, where
$\Omega > \omega_0$.
Then the full phononic Hamiltonian of the system we study becomes
\bea
\H = \hbar \Omega b^\dagger b + \sum_{\k \sigma} \hbar \omega_{\k \sigma} 
b_{\k \sigma}^\dagger b_{\k \sigma} + B u v^2 \, ,
\eea
where $\omega_{\k \sigma}$ are the frequencies of the sample phonons with 
wave vector $\k$ and polarization vector ${\bf e}_\sigma$, and $b$ and $b_{\k 
\sigma}$ are the annihilation operators for the phonons at the adsorbed 
atom and the phonons in the sample, respectively.

\section{Numerical Analysis and Discussion}
\begin{figure} 
$ \left. \right. $
     \includegraphics{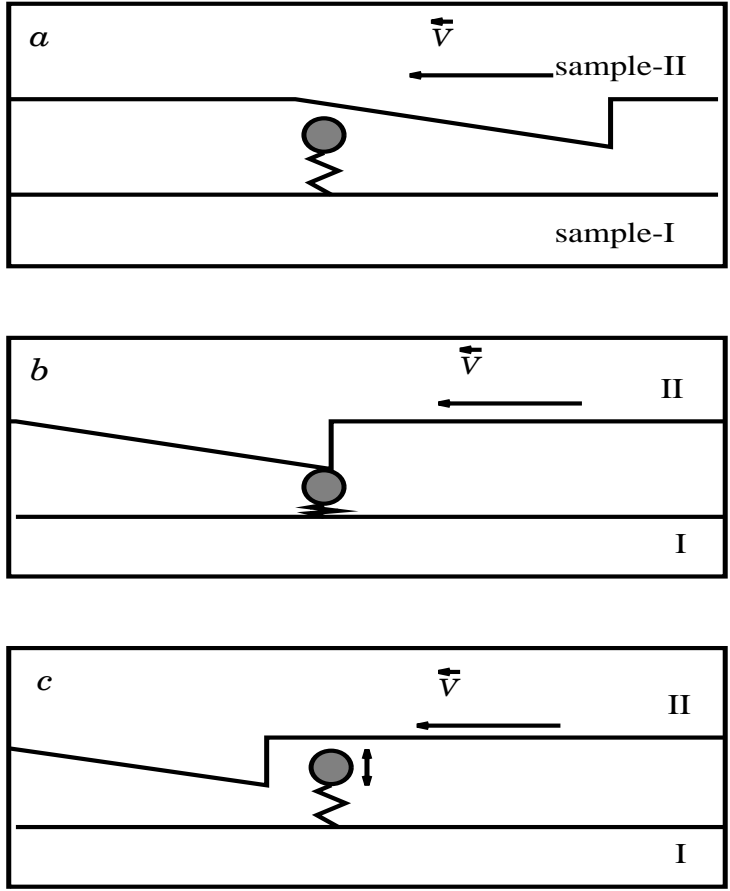}
    \vspace{21.0cm}
\vspace{-8.0cm}
\caption{An adsorbate between the surfaces of two samples, one of which 
moves with a velocity $\vec{v}$. (a) There is no interaction between 
Sample II and the rest of the system. (b) The 
adsorbate is squeezed, absorbing some of the translational energy of 
Sample II. (c) The adsorbate is suddenly released causing it to 
oscillate and the interaction between Sample II and the rest of the 
system can again be neglected.} \label{system}
\end{figure}

In order to construct the initial density matrix, consider the following 
situation: Assume that two samples, Sample I and Sample II, are moving on 
top of one another with 
an adsorbed layer on the bottom one, and there is no direct interaction 
between the samples, as illustrated in Fig. \ref{system}.  
Consider the case when the coverage 
of the adsorbed layer is so low that the interactions between the 
adsorbates can be neglected, in which case one can treat each adsorbate
independently.  During the motion of Sample II, the 
atom adsorbed on Sample I will be pushed and released.  If there 
is a step dislocation on the bottom surface of Sample II, the atom 
will be adiabatically pushed down, due to the wedge shape of the surface, 
displacing it from its equilibrium position and storing energy in it.  
Then it is suddenly released.  After its release there is no interaction 
of the adsorbed atom with Sample II.  This stick-slip model is relevant for 
energy dissipation through phonons in dry sliding friction or 
lubrication, as well as for the vibrations of the adsorbed species.
%(see Fig. 
%\ref{system}). 
The character of the contribution of 
such a mechanism to friction between the bodies would depend on the 
rate of relaxation of this non-equilibrium situation.

Initially, the density matrix of the system plus reservoir is the 
equilibrium density matrix
\bea
\rho = \sum_{\alpha j} \frac{e^{- \beta (\epsilon_\alpha +E_j)}}{Z} \ve 
\alpha j \ra \la \alpha j \ve  \, , 
\eea
where $Z = \sum_{\alpha j} e^{-  \beta (\epsilon_\alpha+ E_j)}$, $\alpha$ 
denotes the number of phonons of the adsorbate and $j$ is a 
multiple index describing the number of phonons in each mode, $\k \sigma$, of 
the sample. 

Adiabatically displacing the atom would not cause the atom to go 
out of equilibrium.  The density matrix will still be diagonal with the same 
diagonal elements but in the displaced basis
\bea
\rho &=& \sum_{\alpha j} \frac{e^{- \beta (\epsilon_\alpha+ E_j)}}{Z} \ve
\alpha' j \ra \la \alpha' j \ve 
\eea
with the same $Z$.  The displaced harmonic oscillator states, $\ve 
\alpha' \ra$, are defined as
\bea
\ve \alpha' \ra = e^{\frac{i}{\hbar} s \hat{p}} \ve \alpha \ra = 
\sum_\beta c_{\alpha' \beta} \ve \beta \ra,
\eea
where $s$ is the displacement of the oscillator and $\hat{p}$ is the 
momentum operator of the adsorbate.
When the adsorbate is suddenly released, the density matrix does not 
change, but now, in the absence of the external force due to  
Sample II, the density matrix is no longer diagonal in the energy eigenstates, 
and the adsorbate is out of equilibrium.  Denoting the RDM of the 
system  right after it is released by $\rho(0^+)$, we have
\bea
\rho(0^+) &=& Tr_b \sum_{\alpha j} \frac{e^{- \beta (\epsilon_\alpha +
E_j)}}{Z} \ve \alpha' j \ra \la \alpha' j \ve \nnb \\
&=& \sum_{\beta \gamma} 
\rho_{\beta \gamma} (0) \ve \beta \ra \la \gamma \ve\,
\eea
where $\rho_{\beta \gamma} (0) = Z^{-1}\sum_{\alpha} c_{\alpha' 
\beta} c_{\alpha' \gamma}^* e^{- \beta \epsilon_\alpha}
$.

Following Ref. [\cite{persson1}], we take
\bea
u &=& \left( \frac{\hbar}{2 m \Omega} \right)^\frac{1}{2} (b + b^\dagger) 
\, , \\
v &=& \sum_{\k \sigma} \left( \frac{\hbar}{2 m \omega_{\k \sigma}N} 
\right)^\frac{1}{2} (b_{\k \sigma} + b^\dagger_{\k \sigma}) {\bf \hat{z} 
.e_{\k \sigma}}.
\eea
where $N$ is the number of the Sample I atoms and ${\bf e}_{\k \sigma}$ is 
the polarization vector of mode $\k \sigma$.  As is pointed out in 
Ref. [\cite{persson1}], 
this expression for $v$ does not account for the surface 
which might reflect bulk phonons, and also does not take into account any 
surface phonons. With these definitions and choosing
\bea
F_1 &=& v^2 \, , \\
Q_1 &=& B u \, ,
\eea
we obtain
\bea
j_{11}(\w) &=& j(\w)=  \theta(\w) \left[ \left(\frac{\hbar}{2 M} \right)^2 
\int d\w' g(\w') g(\w'-\w) 4 \frac{(n_B(\w')+1) 
n_B(\w'-\w)}{\w' (\w' - \w)} + \right. \nnb \\
&+&\left.  \left(\frac{\hbar}{2 M} \right)^2 \int d\w' g(\w') g(\w-\w') 
\frac{n_B(\w')n_B(\w-\w')}{\w'(\w-\w')} \right] + \nnb \\
&+& \theta(-\w) \left[ \left(\frac{\hbar}{2 M} \right)^2 \int
d\w' g(\w') g(\w'-\w) 4 \frac{(n_B(\w')+1) n_B(\w'-\w)}{\w' (\w' - \w)} +
\right. \nnb \\
&+& \left. \left(\frac{\hbar}{2 M} \right)^2 \int d\w' g(\w') g(\w-\w')
\frac{n_B(\w')n_B(\w-\w')}{\w'(\w-\w')} \right] \, , \label{jw} 
\eea
where the integration region in each integration is where the 
density of states is nonzero and $\w$ is positive. In this
result we have assumed the thermodynamic limit and neglected
${\cal O}(\frac{1}{N})$ terms. In this study, $g(\w)$ is represented by   
the Debye density of states
\bea
g(\w) = \frac{3 \w^2}{\w_D^3} \theta(1-\frac{\w}{\w_D})
\eea
where $\w_D$ is the Debye frequency, which is $\w_0$, the maximum
frequency of the substrate.
In order to obtain Eq. (\ref{jw}) from Eq. (\ref{last}), the summations over 
states are converted to integrations over energies and the integration 
region is chosen so that only a small energy violation,  
$\Delta \w$, is allowed, which is assumed to satisfy $\Delta \w 
\Delta t = 1$ 
from the energy-time uncertainty relation.  If one compares Eq. (\ref{jw})
with similar results found in the literature (eg. 
Ref. [\cite{redfield}]), there 
is an extra factor of $\pi$ which arises because of the assumption that 
$\Delta t$ 
is large enough so that one can take the limit $t \rightarrow \infty$ in 
certain integrals.  This factor is not related to the formalism but is 
just related to the evaluation of Eq. (\ref{last}).

The final result can be compared with the results in Ref. [\cite{persson1}].  
In Ref. [\cite{persson1}] it is assumed 
from the beginning that only the diagonal 
element of the density matrix corresponding to the first excited 
state, $\rho_{11}$, is non-zero.  In this case, the contribution of 
the other elements of the density matrix can be neglected in the 
evolution of $\rho_{11}$, and we obtain
\bea
\rho_{11}(t+\Delta t) = \rho_{11}(t) + R_{11;11}(\Delta t) \rho_{11} \, ,
\eea
which yields a decay rate $-\frac{R_{11;11}}{\Delta t}$, which is nothing 
but the result derived in Ref. [\cite{persson1}] 
using the Golden Rule (other than
an overall factor of $\pi$, as discussed earlier).  This 
feature is quite general in the sense that as long as just the first few 
elements of the density matrix are important, and for sufficiently low 
temperatures, the results obtained using this formalism and those 
obtained by the Golden Rule are almost identical.

For the numerical data, we have used the following values:
$\hbar \w_D = 37.6 \, \mbox{meV;\, \,} M=28 \, \mbox{amu;\, \,} m = 28 
\,\mbox{amu;}$ $\hbar \Omega_0 = 46 \, \mbox{meV; \, \,} E_0 = 1.8 \,  
\mbox{eV; \, \,} F = 10^{-10} \,  \mbox{N}$ and $T = 300 \, \mbox{K.}$
Here $F$ is the maximum vertical force applied to the adsorbed atom and 
is related to the vertical displacement, $s$, through
\bea
s = \frac{F}{m \Omega^2} \, .
\eea
We vary the adsorbate frequency $\Omega$ from
$0.2 \Omega_0$ up to $1.52 \Omega_0$, and take the 
iteration step to be $\Delta t = \Omega^{-1}$.

\begin{figure} 
$ \left. \right. $
\vspace{2cm}
     \includegraphics{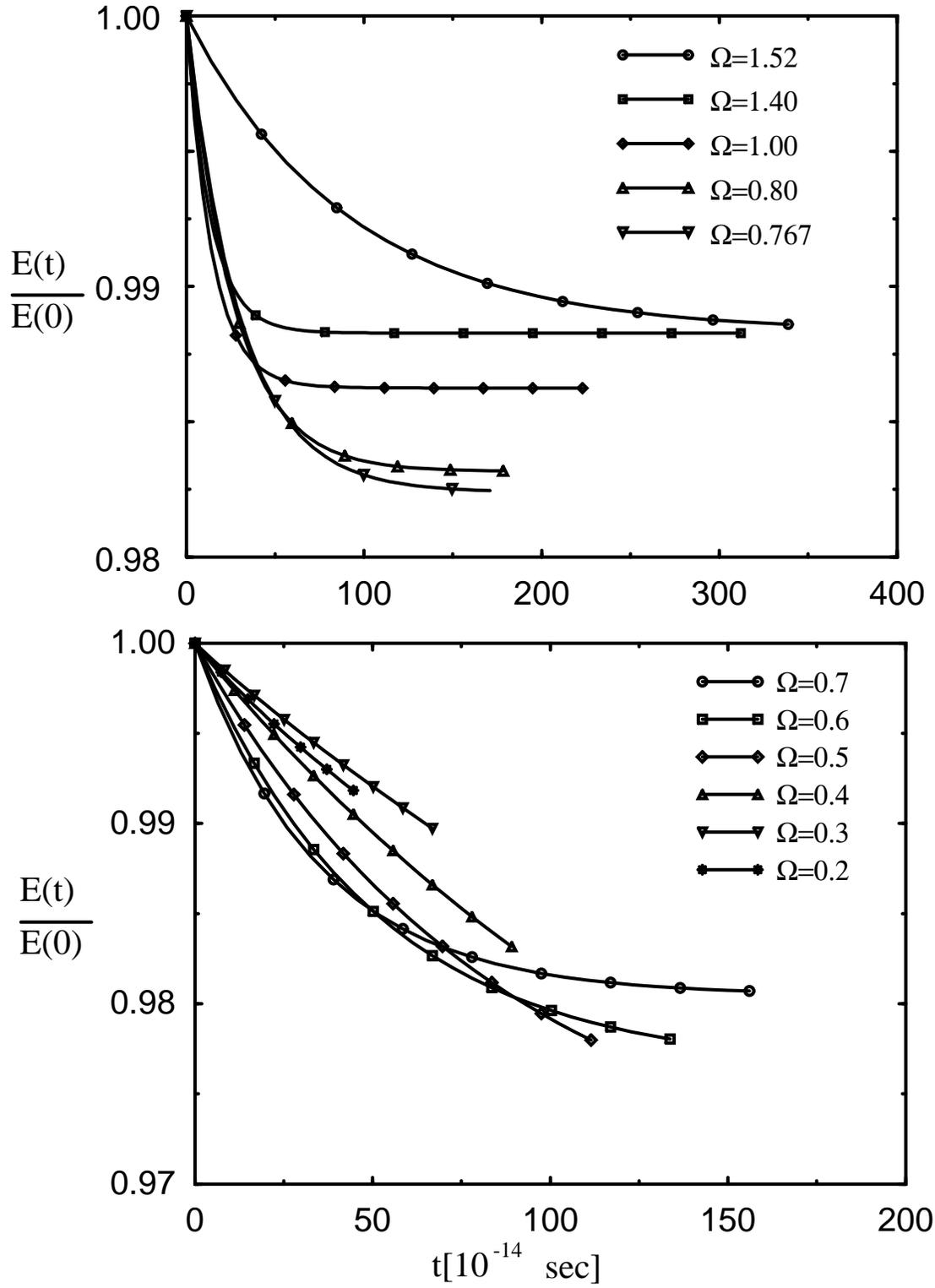}
\vspace{17.5cm}
\caption{Calculated decay profiles, $\frac{E(t)}{E(0)}$, for  the energy of 
the vibrating atom for various $\Omega$'s.}
\label{et}
\end{figure}

In Fig. \ref{et}, we have plotted the decay profiles, 
$\frac{E(t)}{E(0)}$, i.e., the time-dependent energy in the adsorbate
mode over its initial energy,
for various values of the adsorbate frequency, $\Omega$. In each case the range 
of the time axis corresponds to 300 iterations, each iteration 
corresponding to a time of $\Omega^{-1}$.  The exponential character 
of the decay is clear. 
For the numerical calculations, we used only a finite, $16 
\times 16$, part of the infinite density matrix. This caused the matrix 
elements at the edges to evolve incorrectly.  However, as long as they are 
negligible compared to the matrix elements corresponding to the first few 
excited states, this does not affect the general profile of the time 
dependence of the energy, which is mainly determined by the evolution of 
the first few diagonal elements of the density matrix. In most cases 
after 300 iterations the matrix elements at the edge become 
nonnegligible. In all cases, we found that the excess energy can be fit 
almost perfectly to the expression 
\bea
\Delta E(t) = \Delta E(0) e^{-\frac{t}{\tau}},
\eea
where $\Delta E(t) = E(t) - E(\infty)$.
Here, $\tau$ is the decay time constant (or the relaxation time).

\begin{figure} 
$ \left. \right. $
\vspace{16cm}
     \includegraphics{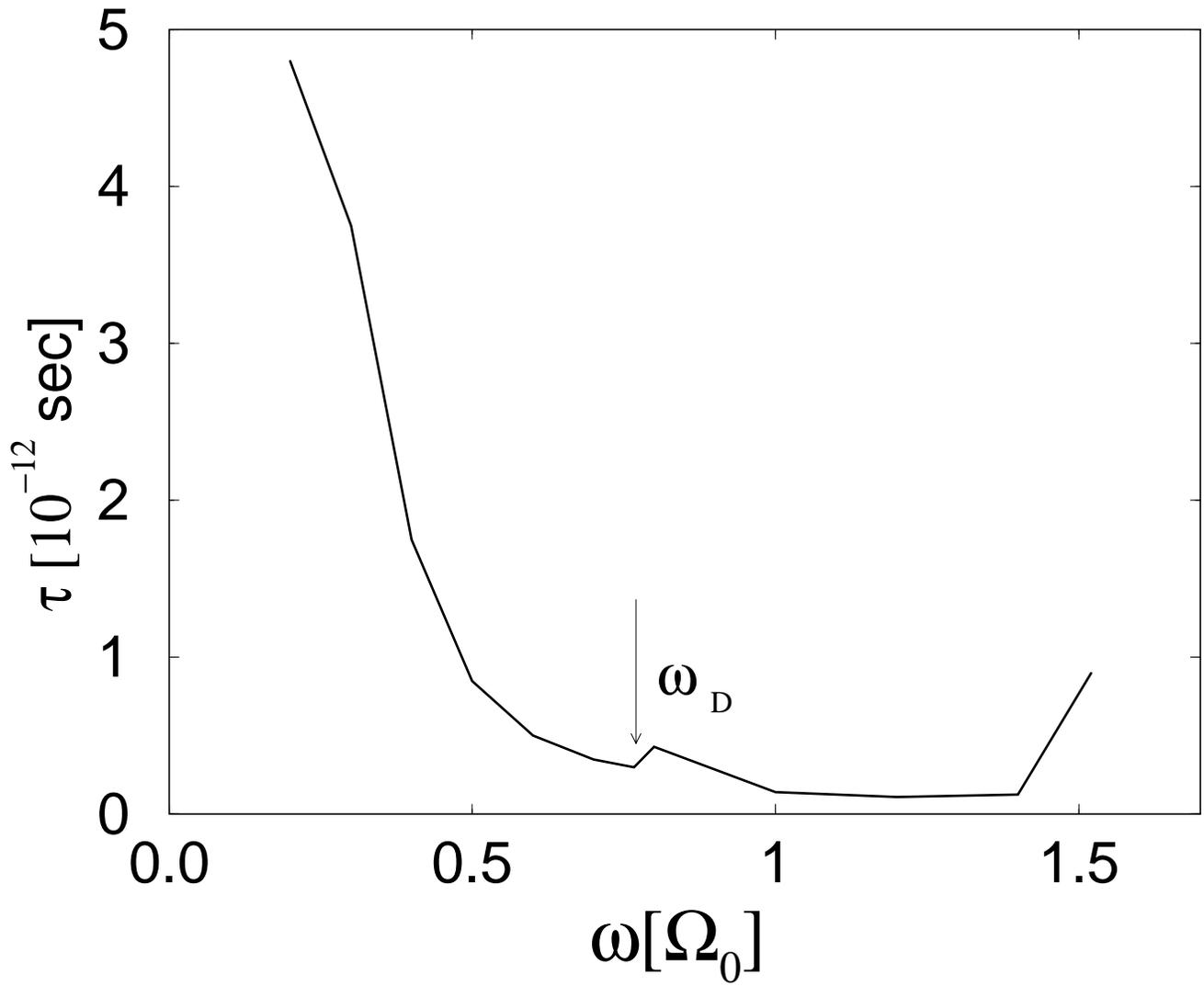}
\caption{Dependence of the relaxation time, $\tau$, on the 
vibrational frequency, $\Omega$, of the adsorbate.} \label{td} 
\end{figure}

In Fig. \ref{td}, the dependence of the decay time, $\tau$, on 
the adsorbate frequency, $\Omega$, is shown.
In the graph, the frequencies are given in units of 
$\Omega_0$, the representative value of the CO-Cu frequency given above.  
We note that our calculations only include cubic anharmonic coupling
of the adsorbate to the substrate.  This is reasonable when $\Omega>\omega_D$. 
When the adsorbate frequency is smaller than the Debye frequency, indicated
in the figure, the plotted $\tau$ will be too large, since we have neglected
linear coupling terms, which are important when $\Omega<\omega_D$.
We nevertheless include the regime where $\Omega<\omega_D$ to show how
the cubic anharmonic contribution to $\tau$ varies in this regime, though
the focus of this study is where the adsorbate frequency lies outside
the band of substrate frequencies.

We see that both for large $\Omega$ and small $\Omega$, 
$\tau$ diverges.  In the large $\Omega$ limit, the reason is due to the phase 
space factors; the two phonons created or absorbed have to be in a band of 
width $2 \w_D - \Omega$ which goes to zero as $\Omega \rightarrow 2 
\w_D$.  For $\Omega > 2 \w_D$, the adsorbed atom cannot decay through 
the emission of two phonons and one has to consider three or more phonon 
processes.  In the small $\Omega$ limit, the coupling constant $B$ 
becomes very small and the system behaves almost as if it is isolated, 
and cannot decay.  In Fig. \ref{td}, one also sees that in the region 
$\Omega \simeq \w_D $ there is an interesting variation in $\tau$.  For $\Omega < 
\w_D$ there is a contribution to the decay process whereby the adsorbate 
absorbs a low energy phonon from the sample which unites with an adsorbate
phonon to emit a 
high energy phonon into the sample.  This process is absent when
$\Omega > \w_D$, which causes $\tau$ to rise when $\Omega$ exceeds $\omega_D$.
We note that this rise, followed by a drop in $\tau$ with larger $\Omega$, is not
affected by the neglect of linear coupling terms since this interesting
nonmonotonic variation of $\tau$ with $\Omega$ occurs where the adsorbate
frequency is higher than the band of substrate frequencies.

\section{Concluding Remarks}
In this work, we developed a formalism based on the reduced density matrix
to study the dissipation of excess energy of an adsorbate into a substrate
in the framework of a stick-slip model for dry friction.  Our focus was on
adsorbate frequencies that lie outside the band of substrate frequencies.  
In this case, anharmonic coupling between the adsorbate and the substrate
controls the rate of energy dissipation, and we introduced cubic terms
explicitly in our calculations. 
This work presents an extension
and generalization of earlier work in which a Golden Rule formula was used
to calculate the rate of energy dissipation from an adsorbate
following the slip step.\cite{blc99}  In that
work, the adsorbate was linearly coupled
to the substrate, and anharmonic terms were not considered.  
In this article, we found that excess energy decays essentially exponentially
from the adsorbate into the substrate, using the reduced
density matrix formalism with anharmonic coupling and
a range of parameters that include those
for CO adsorbed on a Cu substrate at room temperature. 
Previously, we assumed exponential
decay in adopting the Golden Rule.  Here, we have developed a more general formalism
that yields exponential decay for the systems we have investigated at room temperature.

Our work provides an atomic scale theory explaining how and how fast energy dissipates
in a stick-slip model for friction when adsorbates are both small and sparsely
distributed.  When the adsorbate is small, the adsorbate frequency may lie outside the
band of substrate frequencies, and the anharmonic terms that we explicitly 
take into account here will lead to energy dissipation.
Because the adsorbates are distributed sparsely over the substrate, 
we can neglect the interactions among adsorbates in addressing the energy
dissipation from each one.  The model
we have presented fits very well to friction force microscopy.
Our work complements a number of important recent studies that address
energy dissipation by phonons through films of lubricants\cite{csr94,js90,tswk97}
and through the collective motion of asperities,\cite{fs99} systems
where the interactions of the physisorbed or chemisorbed adsorbates give
rise to collective excitations involved in friction.  The theory
we have presented provides a clear picture of phononic energy dissipation
through a nanoparticle, and allows quantitative analysis on this scale.

\newpage

\newpage

\newpage

\end{document}